\documentclass[a4paper,11pt]{article}
\usepackage{jcappub} 

\arxivnumber{2310.15216} 
\title{\boldmath Emergent particles of de Sitter: thermal interpretation of the stochastic formalism and beyond}

\author{TaeHun Kim}
\affiliation{School of Physics, Korea Institute for Advanced Study, \\85 Hoegi-ro, Dongdaemun-gu, Seoul 02455, Republic of Korea}

\emailAdd{gimthcha@kias.re.kr}

\abstract{A thermal interpretation of the stochastic formalism of a slow-rolling scalar field in de Sitter (dS) is given. We construct a correspondence between Hubble patches of dS and particles living in another space called an abstract space. By assuming a dual description of scalar fields and classical mechanics in the abstract space, we show that the stochastic evolution of the infrared part of the field is equivalent to the Brownian motion in the abstract space filled with a heat bath of massless particles. The 1st slow-roll condition and the Hubble expansion are also reinterpreted in the abstract space as the speed of light and a transfer of conserved energy, respectively. Inspired by this, we sketch quantum emergent particles, which may realize the Hubble expansion by an exponential particle production. This gives another meaning of dS entropy as entropy per Hubble volume.}

\begin{document}
\maketitle
\flushbottom

\section{Introduction} \label{sec:introduction}
The de Sitter (dS) space is perhaps one of the most widely studied spacetimes in high energy physics. Being a cosmological solution of the Einstein field equations, the cosmic inflation and the discovery of dark energy show that it approximates our universe in the very early stage and the near future. Local observers in dS experience a cosmological horizon due to a constant expansion rate, which induces various non-trivial behaviors of quantum fields and particles in the observers' frame. It is thus not a surprise that the dS space has received much interest in both practical and theoretical aspects.

Among many other characters, the horizon has drawn much attention within the past decades. A horizon, either cosmological or not, makes a local observer inaccessible to the spacetime region beyond it, and it is now widely accepted that it satisfies thermodynamic relations with quantities from microscopic origins. The first discovery was made for black holes \cite{Bekenstein:1973ur, Bardeen:1973gs}, where a local observer outside the event horizon sees thermal radiation coming from it \cite{Hawking:1975vcx}. Soon realized is that a local observer in dS sees analogous radiation from the cosmological horizon \cite{Gibbons:1977mu}, with the so-called dS temperature $T_{\rm dS} = H/2\pi$ for a Hubble rate $H$. This then opened the area of dS thermodynamics, where the microscopic origins of the thermodynamic quantities are still in debate.

Meanwhile, a similar quantity appears for stochastic dynamics of coarse-grained slow-rolling scalar fields (`infrared (IR) scalar fields') in a dS universe (referring to the flat slicing of a dS space in the present paper) \cite{Starobinsky:1986fx, Starobinsky:1994bd}. Continuously entering fluctuation modes to the superhorizon regime induces a classical stochastic evolution of IR scalar fields subject to a Gaussian random noise of size $\sim T_{\rm dS}$ per Hubble time. The energy density of these fluctuations has a scale-invariant spectrum of $d\rho / d \ln k \sim T_{\rm dS}^4$ at the horizon crossing, and results in the total energy density of $V \sim T_{\rm dS}^4$ after reaching the equilibrium by the potential. These phenomena have been widely studied and applied for inflaton \cite{Clesse:2010iz, Martin:2011ib, PerreaultLevasseur:2013eno, Ramos:2013nsa, Fujita:2013cna, Fujita:2014tja, Vennin:2015hra, Kawasaki:2015ppx, Assadullahi:2016gkk, Barenboim:2016mmw, Vennin:2016wnk, Tada:2016pmk, Pattison:2017mbe, Noorbala:2018zlv, Rudelius:2019cfh, Kitajima:2019ibn, Figueroa:2020jkf, Ando:2020fjm, Figueroa:2021zah, Tada:2021zzj, Pattison:2021oen, Rigopoulos:2021nhv, Ezquiaga:2022qpw, Animali:2022otk, Briaud:2023eae, Tada:2023fvd, Hong:2023pcg} and spectator fields \cite{Finelli:2010sh, Enqvist:2014bua, Hook:2014uia, Hook:2015foa, Espinosa:2015qea, Kohri:2016wof, East:2016anr, Gong:2017mwt, Hardwick:2017fjo, Wu:2020pej, Giudice:2021viw, Jung:2021cps, Graham:2015rva, Arvanitaki:2021qlj, Graham:2018jyp, Fumagalli:2019ohr, Markkanen:2019kpv, Adshead:2020ijf, Nakagawa:2020eeg, Reig:2021ipa, Chatrchyan:2022pcb, Chatrchyan:2022dpy, Takahashi:2018tdu, Ebadi:2023xhq}, as well as in formal aspects and connections to field theoretic approaches \cite{Finelli:2008zg, Gratton:2010zv, Weenink:2011dd, PerreaultLevasseur:2013kfq, Garbrecht:2013coa, Garbrecht:2014dca, Burgess:2014eoa, Burgess:2015ajz, Onemli:2015pma, Karakaya:2017evp, Grain:2017dqa, Collins:2017haz, Pinol:2018euk, Pattison:2019hef, Prokopec:2019srf, Pinol:2020cdp, Cohen:2021fzf, Cruces:2021iwq, Honda:2023unh}; we give only some of the recent examples.

While the stochastic dynamics of IR scalar fields is often intuitively understood by the dS temperature, it is not a thermal effect associated with the cosmological horizon. The stochastic description holds for fields in the superhorizon limit instead of at the horizon scale, and the spin dependence implies that it is not a universal thermal effect arising from the spacetime structure \cite{Finelli:2008zg}. Also, the resultant scale-invariant spectrum is far from thermal distribution, making it hard to obtain a thermal interpretation \cite{Rigopoulos:2013exa, Rigopoulos:2016oko} (c.f. thermal spectrum of gravitationally produced particles for past-asymptotically flat spacetimes \cite{Ford:2021syk, parker1976thermal, hu1996thermal, Koks:1997rk, mersini1998thermal, Biswas:2002qy}). 

Despite these differences, \cite{Rigopoulos:2013exa, Rigopoulos:2016oko} established a thermal interpretation of the stochastic dynamics. The effective action of an IR scalar field is shown to induce a stochastic force that satisfies the fluctuation-dissipation relation with the Hubble friction at $T_{\rm dS}$. This then interprets the stochastic field evolution as a Brownian motion in a medium at that temperature.

Then we ask: can we start with physical observations of dS and arrive at the same conclusion? Here we do this by constructing a correspondence between Hubble patches (region of the 3-dimensional (3D) space of dS covered by a causal patch) and particles living in another space called an abstract space. We propose a dual description of scalar fields and assume classical mechanics in the abstract space. Then, the stochastic evolution of the field turns out to be equivalent to the classical Brownian motion in the abstract space filled with a heat bath of massless particles at $T_{\rm dS}$. Intriguingly, our formalism and model also consistently reinterpret the 1st slow-roll condition as the `speed of light' and the Hubble expansion as a transfer of conserved energy. This gives another meaning for dS entropy, as entropy per Hubble volume in the global dS universe. All these imply that the formalism and the model are more than mere mathematical manipulations.

The paper is composed of three large parts. The first, Secs.~\ref{sec:overview}--\ref{sec:absspacevar}, is a general formalism that we call the emergent particle formalism. The second, Secs.~\ref{sec:heatbathmodel}--\ref{sec:discussion}, is a heat bath model, which applies to the abstract space introduced in the formalism. These two are classical parts, and the third part, Sec.~\ref{sec:quantum}, is a conjecture for the quantum version of the two. 

We begin with the motivation and the setup in Sec.~\ref{sec:overview}, introducing the core concepts of our work. Then in Sec.~\ref{sec:absspacevar}, physical variables in the abstract space are quantitatively identified in terms of ordinary field variables. With these relations, we observe in Sec.~\ref{sec:heatbathmodel} that the stochastic evolution of an IR scalar field is equivalent to the classical Brownian motion in the abstract space, and construct a specific model of heat bath that realizes it through a kinetic theory. In Sec.~\ref{sec:discussion}, we discuss the model parameters and the properties of the abstract space and their further meanings. Being supported by one of them, we further explore and conjecture about quantum emergent particles in Sec.~\ref{sec:quantum}. We then conclude in Sec.~\ref{sec:conclusion}.

The paper uses the natural unit system of $c = \hbar = k_B = 1$. From Sec.~\ref{sec:absspacevar}, the letter $c$ will be used for the speed of light in the abstract space, which differs from the ordinary one.


\section{Motivation and setup} \label{sec:overview}

\begin{figure*} \centering
\includegraphics[width=0.98\linewidth]{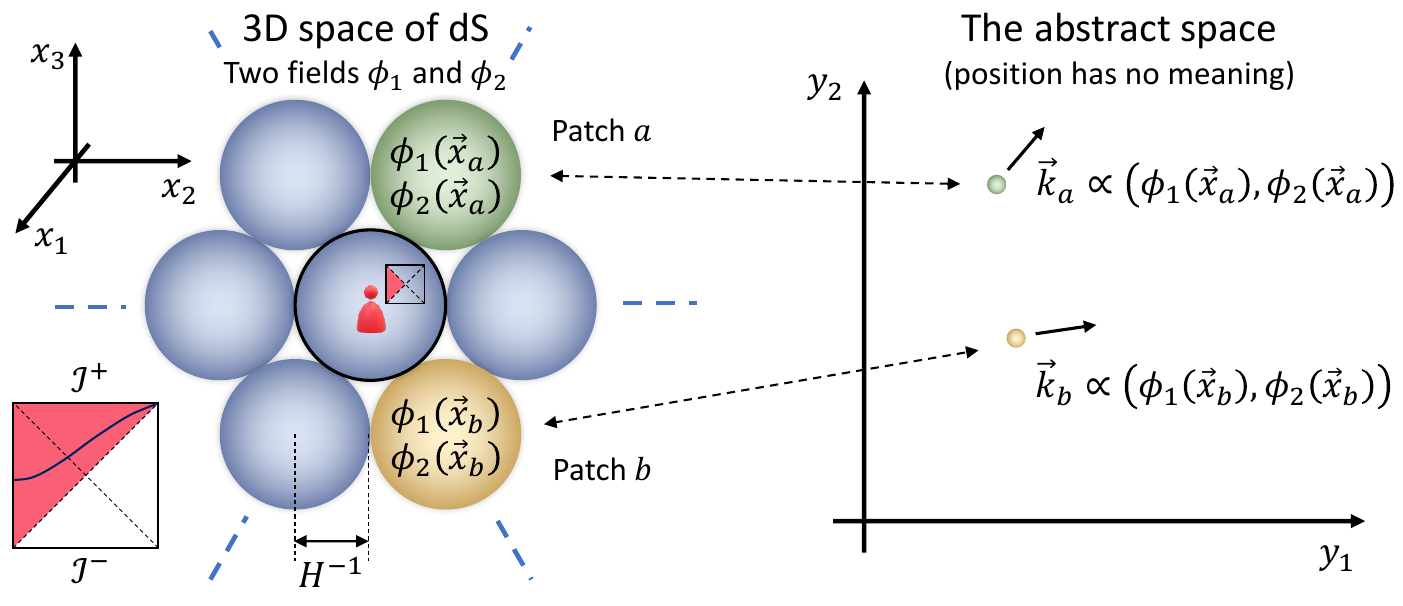}
\caption{A schematic drawing for the emergent particle formalism, for a case of two scalar fields. Left: a constant time slice of a dS universe (a flat slicing drawn as a blue curve in the Penrose diagram) depicted by the emergent particle formalism. Two coarse-grained IR scalar fields $\phi_1$ and $\phi_2$ take different values for different patches (spatial correlation is neglected in drawing). Right: the abstract space with two emergent particles corresponding to the two selected patches (green and yellow). Their momenta $k$'s are given by the field values of $\phi_1$ and $\phi_2$, through Eq.~\eqref{eq:k}.
} 
\label{fig:schematic}
\end{figure*}

The emergent particle formalism depicts time slices of a dS universe to be composed of many horizon-sized Hubble patches. While the horizon is associated with a local observer as a range of causal connections, here we imagine the Hubble patches as `emergent' building blocks of the 3D space in an objective manner. Since any local interaction cannot go over the horizon, the horizon would be a natural size of such quanta if they exist. However, as we will focus on the superhorizon dynamics of scalar fields, the stochastic formalism, this size should be understood as a matter of choice in our work; we are simply using a physically good-looking option. We will also check how our result varies if we choose a different size. 

Under this picture, these patches share similarities with quantum particles, all being mutually equivalent and `particle creation' happens by the Hubble expansion. We call them \textit{emergent particles}, where we use different terminology from quasiparticles as the latter typically arises from microscopic physics. The emergent particles are pieces of the 3D space by themselves, and the virtual space they live in will be called \textit{the abstract space}. This is where the dynamics of emergent particles takes place, and we assume that the usual time variable is shared in the two spaces. Although the picture that the Hubble patches are created violates the energy conservation, we will recover the conservation in the abstract space.

Our picture of dS universe mimics the basic concepts of emergent spacetime theories about (sub-) Planckian hidden structures. However, our building block is macroscopic and not even a physical entity. Hence we emphasize that we are not claiming a new theory, but a formalism that translates several superhorizon aspects of dS into classical and statistical mechanics and thermodynamics in the abstract space.

We infer our proposal for properties of the abstract space by paralleling our situation to basic quantum field theory, aligned with the above similarity. If we consider a dS universe with $n$ independent scalar fields, each field varies with spatial locations of the patches, becoming $n$ continuous degrees of freedom in the abstract space that can differ among the emergent particles. This suggests that the abstract space is $n$D and the field values of patches become an $n$D label of emergent particles, either position or momentum. The former means that the abstract space is the usual field space, whereas we take the latter option, being a \textit{dual description} of the former. This is the first of the two assumed principles in the abstract space. In the rest of the paper, we work with the `minimal non-minimal' setup, a dS universe with one real slow-rolling spectator scalar field $\phi$. The abstract space is then 1D, and we stick to this setup and leave any generalization to future work. 

Since the field is a quantum operator, the correspondence to the `c-number' label is elusive. That being said, the fluctuation modes in the superhorizon scales become indistinguishable from a classical random field \cite{Polarski:1995jg, Lesgourgues:1996jc, Kiefer:1998jk, Kiefer:2008ku}, and each patch can be thought to acquire a classical value for them. It is this IR field that becomes the label of emergent particles in the abstract space. This choice would be reasonable considering both the classicality of the IR field and its homogeneity inside each patch. Hereafter, we deal with the IR field only and denote it with $\phi$. Our dS universe is then classical, so we also seek a model of classical system in the abstract space that can have a correspondence. However, we will see a clue for the quantum version when we complete the model.

To sum up, we summarize the emergent particle formalism in a schematic drawing in Fig.~\ref{fig:schematic}. A Hubble patch in dS with a certain value of an IR scalar field $\phi$ is a particle in the abstract space having the corresponding momentum. $\phi$ becomes a mechanical variable in the abstract space. The set of mutual relations between the physical quantities in the abstract space constitutes classical mechanics in there, and we take \textit{its structure to be identical to the usual classical mechanics}. This is the second of the two assumed principles in the abstract space. Here, the word classical means non-quantum, covering both Newtonian and relativistic regimes. We discuss more on this in Sec.~\ref{sec:absspacevar}.

In the next section, we make correspondences between field variables in dS (`dS field variables') and mechanical variables in the abstract space (`the abstract space variables') from the above two principles aided with heuristic arguments. Before moving on, we state the dS field variables in our setup. The dS universe has a constant expansion rate $H$, arising from an unspecified background energy density $V_0 = 3 M_P^2 H^2$. We have one real minimally coupled slow-rolling spectator scalar field $\phi$ with potential $V_\phi(\phi) \ll V_0$. With loss of some generality, we only consider the cases with $V_\phi$ bounded from below, with nonzero mass at the global minimum. Being bounded is already included in $\phi$ being a spectator field, so only the latter is an additional assumption. Then without loss of generality, we let the minimum point to be at $\phi = 0$, with $V_\phi(0) = 0$, by transferring any constant offset to $V_0$. Near $\phi = 0$, the potential always takes the form of $V_\phi \simeq \frac{1}{2} m_\phi^2 \phi^2$ with $m_\phi^2 \equiv V''_\phi(0) \ll H^2$. Except for these constraints, we let the full potential be an arbitrary function of $\phi$, where our formalism and model can consistently be applied.


\section{The abstract space variables} \label{sec:absspacevar}
We establish the quantitative relations between the abstract space variables and the dS field variables. We work with classical emergent particles until we sketch the quantum version in Sec.~\ref{sec:quantum}. We need the mass $M$, momentum $k$, velocity $v$, and kinetic energy $E_k$ of an emergent particle. We obtain expressions for these in terms of dS field variables, $H$, $\phi$, $V_\phi$, and $m_\phi$, through the assumed dual description and the classical mechanics in the abstract space. 

We start with $E_k$. It is the energy of the emergent particle that depends on the momentum $k$. Since $k$ is associated with $\phi$ through the dual description, we infer that $V_\phi$, the energy density that depends on $\phi$, should be associated with $E_k$. But as $V_\phi$ is an energy density, a volume factor is needed for the conversion. Considering that emergent particles are Hubble patches in 3D space, we choose to use a volume integration over the Hubble volume $4\pi/3H^3$. Therefore, we identify the kinetic energy of an emergent particle as
\begin{equation}
    E_k \, = \, \frac{4\pi}{3H^3} V_\phi, \label{eq:Ek}
\end{equation}
equal to the total potential energy of $\phi$ contained in a Hubble volume. This also makes the equilibrium distribution of $\phi$ \cite{Starobinsky:1994bd}
\begin{equation}
    \rho(\phi) \, \propto \, e^{-\frac{8\pi^2}{3H^4} V_\phi(\phi)} \label{eq:rhophiEquil}
\end{equation}
to be in the form of $e^{-\beta E_k}$ for $\beta = 1/ T_{\rm dS}$. This is a thermal equilibrium distribution of free classical particles in 1D at $T_{\rm dS}$ \cite{Rigopoulos:2013exa, Rigopoulos:2016oko}. Later, the heat bath of our model will be shown to be at $T_{\rm dS}$ too, hence physically consistent by being in a thermal equilibrium. We will revisit this point in Sec.~\ref{ssec:temperature}.

Similarly, $M$ is the energy (see the footnote at the beginning of Sec.~\ref{ssec:classicalheatbath}) of the emergent particle that does not depend on $k$, so it would be associated with $V_0$, the energy density that does not rely on $\phi$. We again use the same volume factor for conversion, resulting in 
\begin{equation} 
    M \, = \, \frac{4\pi M_P^2}{H}, \label{eq:M}
\end{equation}
equal to the total background energy contained in a Hubble volume. Being a spectator field means $V_0 \gg V_\phi$, which implies that the emergent particles are `non-relativistic' in the abstract space, in terms of energy. We comment more on this later in this section. 

Up to this point, we relied on heuristic arguments. We now look at $k$ and $v$ by using the classical mechanics in the abstract space. Their approximating forms near $\phi = 0$ are obtained by $V_\phi \simeq \frac{1}{2}m_\phi^2 \phi^2$. This is where $V_\phi \propto \phi^2$, which corresponds to the Newtonian regime in the abstract space since it means $E_k \propto k^2$ by Eq.~\eqref{eq:Ek} and the dual description. If we rewrite Eq.~\eqref{eq:Ek} by factoring out $M$ in Eq.~\eqref{eq:M}, we get
\begin{equation}
    E_k \, \simeq \, \frac{1}{2 M} \left( \frac{4\pi M_P m_\phi}{\sqrt{3}H^2} \phi \right)^2 \, = \, \frac{1}{2} M \left( \frac{m_\phi}{\sqrt{3} M_P H} \phi \right)^2, \label{eq:Eknearmin}
\end{equation}
giving $k \simeq (4\pi M_P m_\phi / \sqrt{3}H^2) \phi$ and $v \simeq (m_\phi / \sqrt{3} M_P H) \phi$. To obtain the full expressions that are valid also outside the Newtonian regime, we recall the work-energy theorem in classical mechanics in its general form,
\begin{equation}
    E_k \, = \, W \, = \, \int F \, dx \, = \, \int v \, dk . \label{eq:WEtheorem}
\end{equation}
Putting the kinetic energy (\ref{eq:Ek}) in the form of the theorem gives
\begin{equation}
    E_k \, = \, \frac{4\pi}{3H^3} V_\phi (\phi) \, = \, \frac{4\pi}{3H^3} \int^\phi_0 V'_\phi (\phi_1) d\phi_1, \label{eq:Ekint}
\end{equation}
so comparing it with Eq.~(\ref{eq:WEtheorem}) gives that for $k \propto \phi$ following the dual description, we have $v \propto V'_\phi$. From the limiting forms in Eq.~(\ref{eq:Eknearmin}), the full expressions are
\begin{equation}
    k \, = \, \frac{4\pi M_P m_\phi}{\sqrt{3}H^2} \phi \label{eq:k}
\end{equation}
and
\begin{equation}
    v \, = \, \frac{1}{\sqrt{3} M_P H m_\phi} V'_\phi. \label{eq:v}
\end{equation}

Eqs.~(\ref{eq:Ek}), (\ref{eq:M}), (\ref{eq:k}), and (\ref{eq:v}) give the abstract space variables (l.h.s.) in terms of dS field variables (r.h.s.). No ambiguities remain after assuming the volume factor for the conversion between energies in the abstract space and energy densities in our space. These four equations are valid for generic potential $V_\phi$, with the coefficients of Eqs.~\eqref{eq:k} and \eqref{eq:v} fixed by the Newtonian regime near $\phi = 0$. This regime is necessarily the same to ours, by having $E_k \propto k^2 \propto v^2$ arising from $V_\phi \propto \phi^2 \propto V'_\phi{}^2$. Beyond this regime, $V_\phi$ does not satisfy such proportionality and hence $E_k$ too, being analogous to the `relativistic regime'. 

From Eqs.~(\ref{eq:Ek}), (\ref{eq:k}), and (\ref{eq:v}), we see that the functional form of $V_\phi(\phi)$ determines the relations between $E_k$, $k$, and $v$. Therefore, these relations can take generic forms outside the Newtonian regime, making the `relativistic effect' in the abstract space generic too. Each of the relations between the abstract space variables can take own deviations from our relativistic mechanics. For example, there is no requirement that enforces a divergence of $E_k$ ($\propto V_\phi$ by Eq.~\eqref{eq:Ek}) at some finite $v$ ($\propto V'_\phi$ by Eq.~\eqref{eq:v}). The functional forms of $v(k)$ (Eqs.~\eqref{eq:k} and \eqref{eq:v}), $E_k(k)$ (Eqs.~\eqref{eq:Ek} and \eqref{eq:k}), and hence the `relativistic dispersion relation' are fully given by $V_\phi(\phi)$ which we do not have any restriction on its shape. Such a generalized relativistic effect is an inevitable deviation from our physics, allowing us to work with generic potential. However, the Newtonian regime and, much more importantly, the \textit{structure} of the classical mechanics in the whole regime are exactly the same. The latter includes the work-energy theorem, momentum and energy conservation, etc., letting the above equations and the resultant formalism and model be well applied to generic potential. Conversely, being `non-relativistic' in the abstract space has several different meanings. Among them, a slow-rolling spectator field gives non-relativistic emergent particles in the sense that $E_k \ll M$ and $v \ll c$, where $c$ is the `speed of light' in the abstract space (see Secs.~\ref{ssec:classicalheatbath} and \ref{ssec:speedoflight}). Besides these, there is no limitation in the potential shape so other relativistic effects can take place. In App.~\ref{app:SRinabsspace}, we present a concrete example of the abstract space having the same relativistic mechanics as ours. 

Had we used the first integral expression of Eq.~(\ref{eq:WEtheorem}) instead of the second, we would end up with the abstract space being the usual field space. The two descriptions are equivalent, but further physical analogies can be made only by the dual description, translating the field and the potential into mechanical variables. For example, the coefficients in Eqs.~\eqref{eq:k} and \eqref{eq:v} are determined by the Newtonian kinetic energy expression (\ref{eq:Eknearmin}), while there is no such reference if $\phi$ is associated with the position. This dual description leaves no additional fields in the abstract space and all the particles there are free. The only interaction that will be assumed in Sec.~\ref{sec:heatbathmodel} is a contact interaction with another type of particle that enables the collisions between the two. The fact that the emergent particles are free will be used in deriving the momentum statistics for quantum emergent particles in Sec.~\ref{ssec:quantclassdist}. 


\section{Heat bath model of the abstract space} \label{sec:heatbathmodel}
As we established the general formalism in the previous sections, we move on to a specific model that reproduces the stochastic evolution of $\phi$. 


\subsection{The Langevin equation} \label{ssec:Langevin}
The local evolution of slow-rolling IR field $\phi$ is described by the Langevin equation \cite{Starobinsky:1986fx, Starobinsky:1994bd}
\begin{equation}
    d\phi \, = \, -\frac{V'_\phi (\phi)}{3H} dt + \sqrt{\frac{H^3}{4\pi^2}} \, dW \label{eq:Langevin}
\end{equation}
where $dW$ is a unit Gaussian random noise with $\langle dW \rangle = 0$ and $\langle dW^2 \rangle = dt$. The noise comes from the continuous accumulation of stretched IR modes, being Markovian on time and spatial correlation given by the cutoff scale. However, we focus on the local evolution only, which is independent of the cutoff in the IR limit.

Through the emergent particle formalism, we translate Eq.~\eqref{eq:Langevin} into an equation describing a motion of an emergent particle in the abstract space. Substituting Eqs.~\eqref{eq:k} and \eqref{eq:v} into Eq.~\eqref{eq:Langevin} gives
\begin{equation}
    dk \, = \, - \frac{4\pi M_P^2 m_\phi^2}{3H^2} \, v \, dt + \sqrt{\frac{4M_P^2 m_\phi^2}{3H}} \, dW. \label{eq:Langevink}
\end{equation}
By moving to the abstract space, a change of $\phi$ becomes a change of momentum, the slow-rolling term proportional to $-V'_\phi$ becomes a deterministic force proportional to $-v$, and the Gaussian random kicks of $\phi$ become Gaussian random impulses. Since Eqs.~\eqref{eq:k} and \eqref{eq:v} are valid for generic potential, Eq.~\eqref{eq:Langevink} does too. The deterministic force proportional to $-v$ is a familiar form of a drag force. It is smaller for a larger $H$, as momentum conservation in the abstract space corresponds to a static $\phi$ that occurs for $H \rightarrow \infty$. The second term represents the random nature of the momentum changes.

The two terms together mimic the familiar Brownian motion of a particle moving in a medium at a finite temperature. The random collisions provide the random impulses, whereas the directional dependence in the collision rate results in the drag force on average. To see which environment surrounding the emergent particles actually gives the motion in Eq.~\eqref{eq:Langevink}, we think of a model of abstract space filled with a heat bath composed of another type of particle, which we call the bath particle. Bath particles collide with emergent particles and transfer their momentum, obeying the classical mechanics in the abstract space. We present a working model in the following subsection.


\subsection{The classical heat bath model} \label{ssec:classicalheatbath}
The model we present realizes Eq.~(\ref{eq:Langevink}) through a specific type of bath particles. While it is not proven to be unique, finding a proper one is not trivial. We give an example of a failed trial in App.~\ref{app:notworking} and focus on the properly working one here.

The model assumes randomly distributed massless bath particles moving with the same `speed of light' $c \neq 1$ in the abstract space\footnote{We have implicitly normalized the conversion factor between mass and energy to be unity when writing Eq.~\eqref{eq:M}. However, since the relativistic regime in the abstract space is generic, we do not presume the equality between this factor and the square of the speed of light, introduced for massless particles here; see Sec.~\ref{sec:absspacevar}. In other words, we use the normalization of the speed in the abstract space used in Eq.~\eqref{eq:M} in expressing velocities.}, and being absorbed by emergent particles upon collision, similar to familiar cases with photons. We denote the average linear number density of bath particles by $\lambda$ and the bath temperature by $T$. The three parameters, $c$, $\lambda$, and $T$ fully characterize the model. Bath particles with momentum $p$ have energy $c |p|$, and their energy distribution follows the thermal equilibrium distribution $f(E) \propto e^{-\beta E}$ for $\beta = 1/T$. The normalized momentum distribution is then 
\begin{equation}
    f(p) \, = \, \frac{\beta c}{2} \, e^{-\beta c |p|}. \label{eq:fp}
\end{equation} 

We now show that this model reproduces Eq.~(\ref{eq:Langevink}) through the kinetic theory of particles in the abstract space. We first consider momentum conservation at a single collision and then derive statistics for multiple collisions. We do the same with energy conservation in the next subsection. Suppose an emergent particle with momentum $k$ collides with a bath particle with momentum $p$ and absorbs it. Then, its momentum change is simply $\Delta k = p$ by the conservation. In the meanwhile, the average collision rate for bath particles in the momentum range $(p, p + dp)$ is given by
\begin{equation}
    dr(p) \, = \, 
    \begin{cases}
        \lambda (c-v) f(p) dp, & \text{for $p>0$}\\
        \lambda (c+v) f(p) dp, & \text{for $p<0$}
    \end{cases} \label{eq:dr}
\end{equation}
where $v$ is the velocity of the emergent particle, and $|v| < c$ is assumed; we will come back to this point later. 

Then, as outlined in App.~\ref{app:gaussian}, for a long enough (see below) finite time interval $\Delta t$ giving a large number of total collisions, the net impulse $\Delta k$ follows a Gaussian distribution (denoted by $\mathcal{N}(\mu, \sigma^2)$)
\begin{eqnarray}
    \Delta k & \, \sim \, & \mathcal{N} \left( - 2 \lambda v \Delta t \int^\infty_0 p f(p) dp \ , \ 2 \lambda c \Delta t \int^\infty_0 p^2 f(p) dp  \right). \label{eq:Deltakfinal}
\end{eqnarray}
Therefore, we identify the deterministic and the stochastic parts of the momentum change in the form of the Langevin equation as
\begin{eqnarray}
    \Delta k & \, = \, & - \left[2\lambda \int^\infty_0 p f(p) dp \right] v \, \Delta t \, + \, \left[2 \lambda c \int^\infty_0 p^2 f(p) dp \right]^{\frac{1}{2}} \Delta W, \label{eq:DeltakLangevin}
\end{eqnarray}
for a unit Gaussian random noise $\Delta W$ with $\langle \Delta W \rangle = 0$ and $\langle \Delta W^2 \rangle = \Delta t$. This is the same form as Eq.~(\ref{eq:Langevink}), showing that the Brownian motion of an emergent particle due to a heat bath of massless particles is identical to the stochastic evolution of $\phi$ through the emergent particle formalism. The only difference is that $\Delta t$ here is finite, but the required duration to achieve Eq.~\eqref{eq:Deltakfinal} is just a tiny fraction of Hubble time, hence practically indistinguishable; see App.~\ref{app:gaussian}. 

Eq.~\eqref{eq:DeltakLangevin} derived from the heat bath model is valid regardless of the actual relativistic effect of the emergent particle. Clearly, $\Delta k$ at each collision is independent of $v$ and $k$, while only $v$ enters the statistical part by linearly altering the directional dependence of the collision rate in Eq.~\eqref{eq:dr}. These leave no point that actual $v(k)$ can enter. Thus, for generic potential $V_\phi$, the desired stochastic evolution of $\phi$ in Eq.~\eqref{eq:Langevin} is reproduced from this heat bath model, after we fix the three model parameters to match the coefficients of Eqs.~\eqref{eq:DeltakLangevin} and \eqref{eq:Langevink}, and then reverting the variables by Eqs.~\eqref{eq:k} and \eqref{eq:v}. 

To fix the model parameters, we need three equations for them. However, we currently have only two equations coming from the coefficients of Eqs.~(\ref{eq:DeltakLangevin}) and (\ref{eq:Langevink}), so the model is underconstrained. We consider energy conservation and get the third equation in the following subsection.

\subsection{The energy postulation}
Since the collisions between the two types of particles are perfectly inelastic, there must be a loss of kinetic energy for each collision. We first calculate the rate of kinetic energy loss made by continuous collisions per emergent particle. 

From the work-energy theorem (\ref{eq:WEtheorem}), the change of kinetic energy of a massive particle for a small momentum change is $\Delta E_k \simeq v \Delta k$. For a collision between an emergent particle and a bath particle, momentum conservation gives $\Delta k = p$ hence $\Delta E_k \simeq v p$. Meanwhile, the bath particle is absorbed, so the total kinetic energy loss at a single collision is $\Delta E \simeq - |p| c + p v$. Then the average loss rate is obtained by using Eq.~(\ref{eq:dr}) and summing over the two directions as
\begin{eqnarray}
    \frac{\langle \Delta E \rangle}{\Delta t} \, \simeq \, - 2 \lambda (c^2 + v^2) \int^\infty_0 p f(p) dp \, \simeq \, -2 \lambda c^2 \int^\infty_0 p f(p) dp \label{eq:DeltaEfinal}
\end{eqnarray}
in the leading order, with $|v| \ll c$ assumed. This is the average kinetic energy loss rate of the whole abstract space due to each emergent particle. Similar to the momentum case, Eq.~\eqref{eq:DeltaEfinal} holds for generic potential. Eq.~\eqref{eq:WEtheorem} is always valid, and both $\Delta E$ for a single collision and the collision rate depend only on $v$, leaving no dependence on actual $v(k)$.

Since we assumed classical mechanics, we expect the total energy to be conserved in the abstract space so that the lost kinetic energy is converted to other types of energy. If a macroscopic object absorbs photons, the energy excites internal degrees of freedom. But no such internal structure is assumed for the emergent particles. Instead, in the abstract space, the Hubble expansion is a continuous and exponential creation of massive particles, which requires energy gain from somewhere with a rate proportional to the number of existing ones. The required rate of energy gain per emergent particle in the leading order is the particle creation rate times the mass $M$ (we are using $E_k \ll M$ from $V_\phi \ll V_0$ here), giving 
\begin{equation}
    \frac{dE}{dt} \, \simeq \, 3HM \, = \, 12\pi M_P^2. \label{eq:dEdtHub}
\end{equation}

By virtue of energy conservation, we postulate that the kinetic energy loss made by the collisions is converted to emergent particle creation. We call this an \textit{energy postulation}, saying that the Hubble expansion is a transfer of conserved energy in the abstract space. We are not specifying any mechanism, leaving the realization with the quantum emergent particles as a future work. Since Eq.~\eqref{eq:DeltaEfinal} is per emergent particle, the total energy conversion rate is proportional to their number, hence being an exponential production as desired. 

Although the rate in Eq.~\eqref{eq:dEdtHub} is Planckian scale, it does not necessarily render our picture of emergent particle to require full quantum gravity. In the dS universe, this merely reflects the fact that the emergent particles are macroscopic patches (Hubble patches) instead of microscopic (Planckian-sized) regions. The mass $M$ in Eq.~\eqref{eq:M} and the rate in Eq.~\eqref{eq:dEdtHub} are directly proportional to this volume choice (App.~\ref{app:generalderiv}). In the abstract space, the physical constants can differ from our ones and hence the `Planck scale' too. We will obtain $c$ in Eq.~\eqref{eq:c} that differs from unity, while we have no information on $\hbar$ and $G$. A naive assumption that only $c$ differs gives a much greater Planck scale for the abstract space, serving as an example that we should not presume that the Planck scale of the abstract space is the same as ours.

Equating the negative of Eq.~(\ref{eq:DeltaEfinal}) to Eq.~\eqref{eq:dEdtHub} provides the third equation for the model parameters. Then the three parameters are fully determined, and we solve for them in the following subsection.

\subsection{Fixing the model parameters}
From the momentum conservation and the energy postulation, we have
\begin{eqnarray}
    -\frac{\langle \Delta k \rangle}{\Delta t} &\, = \,& 2\lambda v \int^\infty_0 p f(p) dp \, = \, \frac{\lambda v}{\beta c} \, = \, \frac{4\pi M_P^2 m_\phi^2}{3H^2}v \qquad \label{eq:Deltakidentify} \\
    \frac{\sigma^2_{\Delta k}}{\Delta t} &\, = \,& 2\lambda c \int^\infty_0 p^2 f(p) dp \, = \, \frac{2 \lambda}{\beta^2 c} \, = \, \frac{4M_P^2 m_\phi^2}{3H} \label{eq:sigmasqDeltakidentify} \\
    - \frac{\langle \Delta E \rangle}{\Delta t} &\, = \,& 2 \lambda c^2 \int^\infty_0 p f(p) dp \, = \, \frac{\lambda c}{\beta} \, = \, 12\pi M_P^2 \label{eq:DeltaEidentify}
\end{eqnarray}
where $\beta = 1/T$ and Eq.~\eqref{eq:fp} is used. The first two come from Eqs.~\eqref{eq:DeltakLangevin} and \eqref{eq:Langevink} and the last one is from Eqs.~\eqref{eq:DeltaEfinal} and \eqref{eq:dEdtHub}. 

First, $c$ is determined by Eqs.~(\ref{eq:Deltakidentify}) and (\ref{eq:DeltaEidentify}), so dividing the latter by the former gives
\begin{equation}
    c \, = \, \frac{3H}{m_\phi}. \label{eq:c}
\end{equation}
Second, $\beta$ is determined by Eqs.~(\ref{eq:Deltakidentify}) and (\ref{eq:sigmasqDeltakidentify}), so similarly,
\begin{equation}
    T \, = \, \frac{1}{\beta} \, = \, \frac{H}{2\pi} \, = \, T_{\rm dS}. \label{eq:beta}
\end{equation}
And last, $\lambda$ is determined by putting Eqs.~(\ref{eq:c}) and (\ref{eq:beta}) back to any of the three equations,
\begin{equation}
    \lambda \, = \, \frac{8\pi^2 M_P^2 m_\phi}{H^2}. \label{eq:lambda}
\end{equation}
Thus, the heat bath model correctly reproducing the stochastic evolution of $\phi$ as a Brownian motion in the emergent particle formalism is identified with the three model parameters in Eqs.~(\ref{eq:c})--(\ref{eq:lambda}). 


\section{Discussion} \label{sec:discussion}
Hereafter we have several discussions about the formalism and model. After briefly reviewing the formalism, we go through the model in detail.

The emergent particle formalism translated the 3D space of dS into a group of particles. But to describe their dynamics, we introduced another space called the abstract space. This may not seem beneficial, but adopting the dual description of scalar fields made the field variables into mechanical variables. This allowed us to make a direct mechanical analogy of the field's evolution which we arrived at the thermal interpretation through a concrete model of heat bath while keeping the abstract space minimal. All the particles there are free and follow the classical mechanics of the same structure, but allowing generalized relativistic effects determined by $V_\phi(\phi)$ for emergent particles. 

So far we have started suddenly with the notion of emergent particles and adopted the dual description. Had we started only with the latter, we first observe that the Langevin equation describes a stochastic momentum change of some object, while the equilibrium distribution in Eq.~\eqref{eq:rhophiEquil} can be thought as a thermal equilibrium of these objects with kinetic energy proportional to $V_\phi$. Considering that $V_\phi$ is an energy density, we are tempted at this point to interpret those objects as some volume in the 3D space, arriving at the emergent particle formalism. We thus think that the emergent particle and the dual description are closely related in looking at IR scalar fields from a thermal viewpoint.

That being said, the dual description is just a mathematical substitution up to this point. Indeed, any random walking variable can be endowed with a thermal interpretation if we simply declare it to be a momentum in some other space. The original random walking becomes random momentum fluctuations, and then a particle model of the surrounding heat bath may be engineered to reproduce the desired Brownian motion. What gives physical significance to our formalism and model is the consistent reappearance of other physical quantities and phenomena that were not considered in the construction stage. We explore them in the rest of this paper.

\subsection{Bath temperature $T$} \label{ssec:temperature}
We first look at the temperature $T$ of the bath. It determines the momentum distribution of the bath particles and turns out to be the same as $T_{\rm dS}$ as in Eq.~\eqref{eq:beta}. 

The original appearance of the dS temperature in dS space is closely connected with the cosmological horizon of a local observer. In contrast, as mentioned in the introduction, the stochastic formalism of slow-rolling scalar fields concerns the fields in the superhorizon limit. Also, the resulting spectrum does not follow a thermal distribution. 

However, we were able to give a thermal interpretation for the stochastic noise. The quantum-originated Gaussian noise of the field is shown to be equivalent to the Brownian motion of emergent particles in the abstract space, caused by a heat bath at $T_{\rm dS}$. This kind of thermal interpretation is also noticed in \cite{Rigopoulos:2013exa, Rigopoulos:2016oko}, but we arrived at the same conclusion from physical observations and postulations about dS. 

Then what about the spectrum? The stochastic noise appearing in Eq.~\eqref{eq:Langevin} is about the local evolution of the field, while the spectrum is about the field profile in different spatial scales. Therefore, in the language of emergent particles, it is a matter of how the emergent particles are bonded to give a conventional 3D space. This is beyond the scope of our paper and left as a future work. 

The fact that the bath temperature is $T_{\rm dS}$ makes the model physically consistent. Since the model parameter $T$ is the temperature of the heat bath, we expect that the emergent particles will eventually thermalize and reach the same temperature. As discussed with Eq.~\eqref{eq:rhophiEquil}, the equilibrium solution of the Fokker-Planck equation for $\phi$ is already a thermal equilibrium distribution of emergent particles at $T_{\rm dS}$. So the Fokker-Planck equation describes the thermalization process in the abstract space. Indeed, the coefficients of Eq.~\eqref{eq:Langevink} satisfy the fluctuation-dissipation relation at $T_{\rm dS}$, indicating that if the motion arises from a thermal background, it should be at that temperature. However, explicit construction of the particle model of the heat bath goes beyond this, as will be explored below. 

An interesting point is that although the dS temperature and $T$ here have different origins, they coincide in value. To check at which point this arose, we make a most general derivation of the previous results in App.~\ref{app:generalderiv}. As can be seen there, $T$ is proportional to the volume of an emergent particle in our space, which we took to be the Hubble volume. While the equality between the bath temperature and the equilibrium temperature is not affected, the temperature itself is affected. The coincidence between $T_{\rm dS}$ and $T$ happens only if we take the volume of an emergent particle to be the Hubble volume; the horizon. Any possible deeper origin of this coincidence is left to future work.

\subsection{Speed of light $c$} \label{ssec:speedoflight}
The speed of light $c$ in the abstract space is given by Eq.~(\ref{eq:c}). It was introduced as the speed at which the massless bath particles move. Then, as we assumed classical mechanics, we expect it to be the speed limit for the massive particles. This is why we assumed $|v| \ll c$ in the kinetic theory of particles in Sec.~\ref{sec:heatbathmodel}.

Since Eq.~(\ref{eq:v}) converts the field's potential slope to the velocity of an emergent particle, we can revert the velocity in terms of the potential slope. How large is the slope when $v \sim c$ is reached? Is the assumption $|v| \ll c$ an additionally imposed one, or does it have a counterpart in the usual field description?

From Eqs.~(\ref{eq:v}) and (\ref{eq:c}), we have
\begin{equation}
    \left. V'_\phi \right|_{v=c} \, = \, 3\sqrt{3} M_P H^2. \label{eq:cVp}
\end{equation}
Interestingly, this is the value at which the 1st slow-roll condition is violated. The 1st potential slow-roll parameter is $\epsilon_V  \simeq M_P^2 \times (V'_\phi/3 \sqrt{2} M_P^2 H^2 )^2$. Therefore, the potential slope where $\epsilon_V = 1$ happens is $V'_\phi = 3\sqrt{2} M_P H^2$, having only a factor of $\sqrt{2/3}$ difference from Eq.~(\ref{eq:cVp}).

The reappearance of the 1st slow-roll condition is unexpected but consistent with our picture. In the usual field theoretic description, the potential slope for a slow-rolling field is limited to have $\epsilon_V < 1$, and $\epsilon_V = 1$ is where the (quasi-) dS expansion of the background spacetime breaks down. In the abstract space with classical mechanics, if $c$ is the speed of light, we expect it to be the speed limit for the massive particles and where physics in $|v| \ll c$ breaks down. These two correspond to each other by the identification of $v$ in Eq.~\eqref{eq:v}. We leave any analysis of the actual relativistic effect near $v \sim c$ (corresponds to $\epsilon_V \sim 1$) in the abstract space to future work. 

This also supports the energy postulation. As can be seen in the derivation of Eq.~(\ref{eq:beta}), determining $c$ requires information about the energy. Indeed, $c$ is the conversion factor between the momentum and the energy of the bath particles. If, for example, we equate Eq.~\eqref{eq:DeltaEfinal} to some constant multiples of Eq.~\eqref{eq:dEdtHub}, i.e. an additional factor appearing in the r.-most-h.s. of Eq.~\eqref{eq:DeltaEidentify}, the square root of that factor appears in the r.h.s. of Eq.~\eqref{eq:c}; see App.~\ref{app:generalderiv}. So although there is a mild constant factor ambiguity, it is clear that the agreement between $c$ and the 1st slow-roll condition can be made only if the energy loss rate matches (in orders of magnitude) the one required for the Hubble expansion. Hence the agreement again supports our picture, being a result of energy conservation that is expected after the assumed classical mechanics in the abstract space. All these consistencies give our formalism and the model a physical significance, being more than a heat bath engineering. Furthermore, the thermal interpretation is extended to the Hubble expansion, by interpreting it as a particle production due to a transfer of conserved energy from the heat bath to emergent particles.

\subsection{The flat abstract space} \label{ssec:propertyabs}
We discuss more about the properties of the abstract space, the assumed symmetries and their further implications.

The assumption of classical mechanics implies that the abstract space, together with the shared time variable, has time and spatial translation symmetries. The energy and momentum of an emergent particle change only by collisions with bath particles (Eqs.~\eqref{eq:DeltakLangevin} and \eqref{eq:DeltaEfinal}), and are otherwise conserved all the way through its motion. On the other hand, the generalized relativistic effect we discussed in Sec.~\ref{sec:absspacevar} indicates that the abstract space does not have the Lorentz symmetry in general, though its results are restored for a specific potential presented in App.~\ref{app:SRinabsspace}. And last, we have not probed rotational symmetry as we have remained in 1D abstract space.

Therefore, the abstract space (plus time; 1+1D) can be thought of as a flat spacetime but without Lorentz symmetry in general. The emergent particles and the heat bath constitute a thermal system in a flat spacetime. Consequently, the classical-level correspondence\footnote{This is the correspondence between the emergent particles and the stochastic formalism. For the correspondence between the stochastic formalism and the usual quantum field theoretic description, we refer to the literature listed in the introduction.} we saw can be said to be a partial correspondence between IR scalar fields in dS and thermal systems in flat spacetime. We saw the correspondence for the local stochastic evolution of the field, which encodes the field's probability distribution and temporal correlations at a fixed spatial point, and the 1st slow-roll condition, which gave physical significance to our formalism and model. However, we could not reach general correlation functions and other conditions. The spatial information of the correlation will be encoded in the way we place the emergent particles to recover the conventional 3D space, which is beyond the scope of the present paper, as also briefly mentioned in Sec.~\ref{ssec:temperature}. Also, we are not yet clear about the dS counterpart for the bath particles in the abstract space. Searching for these missing counterparts and establishing a more complete correspondence will be interesting future work.

Although the energy postulation remains to be realized as emergent particle production, the correspondence at the quantum level will also cover the Hubble expansion of dS itself. We discuss more on quantum emergent particles in the next section.


\section{Conjecture for quantum emergent particles} \label{sec:quantum}
Since the energy postulation is supported by the agreement between $v \simeq c$ and $\epsilon_V \simeq 1$, we further sketch the quantum emergent particles that would realize the postulation. In a quantum field theoretic description, we will have two types of fields, each for the massive emergent particle and the massless bath particle. They will have a certain kind of interaction, which makes the former absorb energy from the latter. The net process is one way, which is likely to rely on some asymmetry either mathematical or physical. The energy absorption should result in particle production like in many particle cosmology scenarios. 

If we have a successful quantum version, we will have a consistent thermal description for both the Hubble expansion and the (local) evolution of IR scalar fields. In the ordinary description, a constant energy density makes the universe expand with a constant Hubble rate. Quantum fluctuations of scalar fields are stretched and frozen, and their continuous entrance to the IR regime gives the stochastic evolution. In the abstract space, the starting point would be the heat bath filling the abstract space. It induces the thermal motion of emergent particles, and the interaction between the two fields makes an energy transfer and emergent particle creation. The two descriptions are equivalent by the relations between the variables in the two spaces, Eqs.~\eqref{eq:Ek}, \eqref{eq:M}, \eqref{eq:k}, and \eqref{eq:v}. 

We leave detailed model buildings to future work and discuss two features that we anticipate the quantum version to show.

\subsection{Quantum particles with classical distribution} \label{ssec:quantclassdist}
The immediate problem we face is that both boson and fermion have equilibrium distributions of $E_k$ different from the Maxwell-Boltzmann distribution of $\propto e^{-\beta E_k}$. These cannot reproduce the equilibrium distribution of the Fokker-Planck equation in Eq.~\eqref{eq:rhophiEquil}. 

Here we show an example that can detour this problem. The key idea is to effectively revive the distinguishability of quantum emergent particles. First, introduce another space that we call the auxiliary space. Then, attach it to the abstract space and let the emergent particle simultaneously live in both spaces. Now assume that the emergent particle is a boson in the abstract space and a fermion in the auxiliary space. In terms of annihilation operators, we write
\begin{equation}
    \hat{o}_{k, \vec{q}} \, = \,\hat{a}_k \otimes \hat{b}_{\vec{q}} \label{eq:mixptldef}
\end{equation}
where $\hat{o}_{k, \vec{q}}$ is the annihilation operator for the emergent particle, $\hat{a}_k$ is the bosonic annihilation operator in the abstract space, and $\hat{b}_{\vec{q}}$ is the fermionic annihilation operator in the auxiliary space. $k$ is the momentum in the abstract space as above, and $\vec{q}$ is the momentum in the auxiliary space. We use a vector symbol for the latter since the dimensionality of the auxiliary space is not known. $\hat{a}_k$ and $\hat{b}_{\vec{q}}$ satisfy the usual commutation and anticommutation relations, respectively, as
\begin{eqnarray}
    [\hat{a}_{k_1}, \hat{a}_{k_2}^\dagger] \, \propto \, \delta(k_1 - k_2) \ &\text{,}&  \quad [\hat{a}_{k_1}, \hat{a}_{k_2}] \, = \, [\hat{a}_{k_1}^\dagger, \hat{a}_{k_2}^\dagger] \, = \, 0 \label{eq:commboson} \\
    \{\hat{b}_{\vec{q}_1}, \hat{b}_{\vec{q}_2}^\dagger\} \, \propto \, \delta(\vec{q}_1 - \vec{q}_2) \ &\text{,}& \quad \{\hat{b}_{\vec{q}_1}, \hat{b}_{\vec{q}_2}\} \, = \, \{\hat{b}_{\vec{q}_1}^\dagger, \hat{b}_{\vec{q}_2}^\dagger\} \, = \, 0. \label{eq:commfermion}
\end{eqnarray}

Due to the anticommutativity of $\hat{b}_{\vec{q}}$, all the emergent particles must have different $\vec{q}$'s regardless of their $k$'s. For example, a state with two emergent particles with the same $k$ can exist,
\begin{equation}
    \hat{o}^\dagger_{k_1, \vec{q}_1} \hat{o}^\dagger_{k_1, \vec{q}_2} |0\rangle \, = \, \hat{a}^\dagger_{k_1} \hat{a}^\dagger_{k_1} \otimes \hat{b}^\dagger_{\vec{q}_1} \hat{b}^\dagger_{\vec{q}_2} |0 \rangle
\end{equation}
but if they have the same $\vec{q}$, such a state does not exist:
\begin{equation}
    \hat{o}^\dagger_{k_1, \vec{q}_1} \hat{o}^\dagger_{k_2, \vec{q}_1} |0\rangle \, = \, \hat{a}^\dagger_{k_1} \hat{a}^\dagger_{k_2} \otimes \hat{b}^\dagger_{\vec{q}_1} \hat{b}^\dagger_{\vec{q}_1} |0 \rangle \, = \, 0.
\end{equation}
Note that the latter holds regardless of their $k$'s. Emergent particles have exclusiveness on $\vec{q}$ but they can have any occupation number on any $k$ state.

What we want is the statistics of $k$ in the abstract space. We project the two-space statistics onto the abstract space by marginalizing the auxiliary space. Then, since states with the same $k$ occupation numbers but different $\vec{q}$ combinations are counted individually, the exclusiveness on $\vec{q}$ effectively revives the distinguishability of the emergent particles in the abstract space. As a result, the statistics of classical particles can be obtained with quantum particles. We explicitly show this in App.~\ref{app:mixedptl} for free particles.

The introduction of the auxiliary space resembles the slave-boson or slave-fermion (parton in a more general sense) approaches used for strongly correlated electron systems in condensed matter theories \cite{Coleman:1984zz, yoshioka1989slave}. However, these approaches require a joint constraint on the two separated operators (corresponds to $\hat{a}_{k}$ and $\hat{b}_{\vec{q}}$ here) in addition to the canonical (anti)commutation relations of each, to preserve the fermionic property of the original electron operator. Here, no such constraint is employed, and the emergent particle follows the classical statistics. 

An ambiguous point is the counterpart of $\vec{q}$ in dS. An immediate proposal would be the 3D space itself since if we view the 3D space as composed of many Hubble patches, they do not overlap; the exclusiveness. We refrain from making any further speculations and leave all these to future work.

\subsection{Thermodynamics of global de Sitter universe} \label{ssec:globaldSthermodynamics}
The Hubble expansion is expected to be realized as an exponential production of emergent particles. The global dS universe is a quantum field of the emergent particle, and it absorbs energy from the heat bath, and the particles are created. We apply the 1st law of thermodynamics $dQ = TdS - PdV$ to this process. 

After the emergent particles reach the thermal equilibrium, they are at the dS temperature as discussed in Sec.~\ref{ssec:temperature}. Since $M \gg E_k$, approximately for each $M$ amount of energy inflow, one emergent particle is created. The subtle part of the 1st law is the pressure and the volume, but it would be hard to associate any kind of work with the process of particle creation, so we set $PdV = 0$.

Then, for $\Delta Q = M$, the 1st law reads
\begin{equation}
    \Delta Q \, = \, \frac{4\pi M_P^2}{H} \, = \, T \Delta S \, = \, \frac{H}{2\pi} \Delta S
\end{equation}
so 
\begin{equation}
    \Delta S \, = \, \frac{8 \pi^2 M_P^2}{H^2} \, = \, S_{\rm dS}
\end{equation}
for each emergent particle creation. Therefore, the dS entropy reappears and acquires another meaning in the global universe, as entropy per emergent particle or entropy per Hubble volume. Since entropy is an extensive quantity whereas temperature is intrinsic, this definition holds for all the emergent particles at all times.

The thermodynamic process we consider differs from the one considered in local descriptions of dS. There, a change of the horizon associated with an energy flow or cosmological constant variation is considered; see \cite{Padmanabhan:2002sha, Frolov:2002va, Sekiwa:2006qj} for examples. The description of the process and the identification of thermodynamic quantities are made in the usual spacetime. On the other hand, we look at the global dS universe as one system and consider the Hubble expansion as a thermodynamic process. We describe it in the abstract space and infer the vanishing work. 

We comment on the scaling behavior of $\Delta S$ per emergent particle with referring App.~\ref{app:generalderiv}. From the same scaling of $M$ and $T$ with respect to the volume of an emergent particle in our space, $\Delta S$ per emergent particle is not affected by the volume choice. However, this means that when we revert to our space, the entropy per Hubble patch is inversely proportional to the volume choice. So like the case of temperature in Sec.~\ref{ssec:temperature}, the agreement between the two entropies with different origins holds when we take the size of the emergent particle to be the Hubble volume; the horizon again. We leave any further studies on this coincidence for future work.


\section{Conclusion} \label{sec:conclusion}
The stochastic formalism of IR scalar fields shows several similarities with conventional thermodynamics or dS thermodynamics. The sizes of quantum fluctuations, resultant energy density, and the energy spectrum at the horizon crossing are all roughly described by one parameter, $T_{\rm dS}$. The resultant equilibrium distribution mimics the thermal distribution, while the Langevin equation reminds the Brownian motion. That being said, it has not been understood as a thermal effect nor related to the dS thermodynamics where the cosmological horizon plays a pivotal role.

Our work provides a thermal interpretation of the stochastic evolution of IR scalar fields through the emergent particle formalism. The emergent particles living in the abstract space correspond to patches of the 3D space of dS. We implemented the dual description of scalar fields and assumed classical mechanics in the abstract space, and derived the relations between the abstract space variables and dS field variables. Then, we showed that the Langevin equation describing the stochastic evolution of the IR scalar field can be translated into an equation describing the Brownian motion in the abstract space, filled with a heat bath of massless particles. This correspondence holds for generic slow-rolling spectator potential.

The formalism and the model turned out to do more than intended. The 1st slow-roll condition became the speed of light in the abstract space, which is consistent with our assumption of classical mechanics. The formalism relates the potential slope to the velocity in the abstract space, so the limit of the former should be the limit of the latter, which would be the speed of light. The fact that this agreement holds only when the energy lost by inelastic collisions is equated to the energy required for the Hubble expansion supports energy conservation in the abstract space, again being consistent with the assumed classical mechanics. This reinterprets the Hubble expansion as a transfer of conserved energy from the heat bath to emergent particles. We then sketched quantum emergent particles that we hope to realize the Hubble expansion as particle production and saw another meaning of the dS entropy in the global dS universe.

Although interesting results have appeared, our work can operate only after a setup of a dS universe is given. Namely, it cannot explain how the relevant quantities are determined. For example, we cannot determine the mass $M$ of an emergent particle for a given heat bath temperature. On the contrary, the usual gravitational description starts with a constant energy density and this determines everything. In addition to this, we are left with reconstructing the 3D space from emergent particles if we want to recover the spatial information about IR scalar fields in the emergent particle picture.

Despite these shortcomings, our work may serve as a starting point for several directions of future work. For example, on the theoretical side, any deeper relation to the dS thermodynamics arises as an immediate question. While the stochastic formalism is unrelated to the cosmological horizon, we saw that the reappearance of the dS temperature and dS entropy per Hubble patch happens only when we choose the emergent particle to be horizon-sized in our space. On the practical side, one may ask whether our formalism can give an advantage in calculating inflationary quantities after incorporating deviations from the exact dS expansion. Last but not least, the classical correspondence we saw could be a primitive form of a more complete correspondence between a dS universe with IR scalar fields and thermal systems in a flat spacetime. If we successfully fill out the missing counterparts, we may reach a new understanding of the dS scalar field theory.

In all, the emergent particle formalism and the heat bath model open new perspectives on the IR scalar fields in dS and even on the Hubble expansion. We hope our work may shed new light on relevant topics in the future.


\acknowledgments

The author greatly thanks Junghwan Lee, Kiyoharu Kawana, Jinwoo Sung, Seung Hun Lee, Minju Kum, Suro Kim, Dhong Yeon Cheong, Sunghoon Jung, Jinn-Ouk Gong, Wei-Chen Lin, Maria Mylova, Ki-Young Choi, Inyong Cho, Han Gil Choi, Dong Woo Kang, Wan-Il Park, and Seoktae Koh for many helpful discussions and comments. The author is supported by a KIAS Individual Grant PG095201 at Korea Institute for Advanced Study.


\appendix

\section{Example of the relativistic mechanics in the abstract space} \label{app:SRinabsspace}
We give an example of the potential that can reproduce the usual relativistic mechanics in the abstract space. If $V_\phi$ is
\begin{eqnarray}
    V_\phi (\phi) &=& \sqrt{27H^4 M_P^2 \phi^2 + \frac{729H^8 M_P^4}{m_\phi^4}} - \frac{27 H^4 M_P^2}{m_\phi^2} \nonumber \\
    &\simeq& \frac{1}{2} m_\phi^2 \phi^2 - \frac{1}{216} \frac{m_\phi^6}{H^4 M_P^2} \phi^4 + \frac{1}{11664} \frac{m_\phi^{10}}{H^8 M_P^4} \phi^6 + \cdots \label{eq:VphiSR}
\end{eqnarray}
with $m_\phi = 3H$, one can explicitly check that $E_k$, $M$, $k$, and $v$ following from Eqs.~\eqref{eq:Ek}, \eqref{eq:M}, \eqref{eq:k}, and \eqref{eq:v} satisfy the special-relativistic mechanics, with the speed of light $c = 1$. However, since this is not a slow-rolling field, $V_\phi$ in Eq.~\eqref{eq:VphiSR} lies out of the validity range of the Langevin equation~\eqref{eq:Langevin}. Instead, one may set $m_\phi \ll H$ to make $\phi$ a slow-rolling field. This then gives the special-relativistic mechanics but with $c = 3H/m_\phi \neq 1$. This example shows that the correspondence between the classical mechanics in our space and the one in the abstract space can be accurate, depending on the actual potential shape. 


\section{Non-working model of heat bath: massive bath particles} \label{app:notworking}
To demonstrate that the working model in Sec.~\ref{sec:heatbathmodel} is not a trivial one, we present a non-working model of massive bath particles. Due to the extended velocity distribution, we end up with a drag force not proportional to the velocity.

As our purpose is to provide an example, we work with the simplest case. The bath particles with mass $m$ and number density $\lambda$ follow a thermal equilibrium distribution $f(E)\propto e^{-\beta E}$. We assume their massive nature gives $p = m u$ and $E = \frac{p^2}{2m}$ for velocity $u$. The distribution in the velocity domain is 
\begin{equation}
    f(u) \, = \, \sqrt{\frac{\beta m}{2\pi}} e^{-\beta \frac{1}{2} m u^2}. \label{eq:fu}
\end{equation}

The bath particles are assumed to make elastic collisions with emergent particles; we give up accounting for the Hubble expansion and focus only on the momentum changes. For the emergent particle, we focus on a particular case with $V_\phi = \frac{1}{2} m_\phi^2 \phi^2$ for a concrete demonstration. Then, both the emergent particles and the bath particles exactly follow Newtonian mechanics. As a result, a single collision between an emergent particle with velocity $v$ and momentum $k$ and a bath particle with velocity $u$ and momentum $p$ results in the momentum change of the former $\Delta k = \frac{2Mm}{M+m}(u-v)$.

Meanwhile, the collision rate is now a function of $u$. Similar to Eq.~(\ref{eq:dr}), the average collision rate for bath particles in the velocity range $(u, u + du)$ is given by
\begin{equation}
    dr(u) \, = \, \lambda |u-v| f(u) du.
\end{equation}
Then, the average momentum change rate for an emergent particle with velocity $v$ is 
\begin{eqnarray}
    \left\langle \frac{dk}{dt} \right\rangle & \, = \, & \frac{2Mm}{M+m} \lambda \int^\infty_{-\infty} (u-v)|u-v|f(u) du \nonumber \\
    & \, = \, & - \frac{2M}{M+m}  \frac{\lambda}{\beta} \left[ \frac{2}{\sqrt{\pi}} e^{-x^2} x + (1+2x^2) \, \text{erf}(x) \right] \label{eq:dkdtavgnotworking}
\end{eqnarray}
where $x \equiv \sqrt{\beta m / 2} \, v$ and Eq.~(\ref{eq:fu}) is used in evaluating the integral. Eq.~(\ref{eq:dkdtavgnotworking}) is non-linear in $x$, especially when $|x|\gtrsim 1$ that the emergent particle reaches the typical speed of the bath particles. 

Therefore, we conclude that the model with massive bath particles is wrong as it cannot correctly reproduce the drag force proportional to $-v$. Observing that the extended velocity distribution hinders obtaining $\langle dk/dt \rangle \propto -v$ in Eq.~(\ref{eq:dkdtavgnotworking}), we considered the massless bath particles in Sec.~\ref{ssec:classicalheatbath}.


\section{Achieving net Gaussian impulse in a finite time} \label{app:gaussian}
We show how and how fast the net impulse $\Delta k$ follows a Gaussian distribution (\ref{eq:Deltakfinal}) from the average collision rate (\ref{eq:dr}) and the momentum distribution (\ref{eq:fp}). 

For a finite time interval $\Delta t$, integrating Eq.~(\ref{eq:dr}) over $p$ gives on average $\frac{1}{2}\lambda(c-v)\Delta t$ number of collisions for right-moving ($p>0$) bath particles and $\frac{1}{2}\lambda(c+v)\Delta t$ for left-moving ($p<0$) ones. The actual number of collisions for each direction $N^\pm$ is a random variable following the Poisson distribution (denoted by $\text{Pois}(\mu)$)
\begin{equation}
    N^\pm \, \sim \, \text{Pois}\left(\frac{1}{2}\lambda(c \mp v)\Delta t\right). \label{eq:Npmdist}
\end{equation}

Meanwhile, the bath particle speed $c$ is independent of their momentum $p$, so is the average collision rate per number of particles. Then, the total impulse during $\Delta t$ by $N^+$ right-moving particles is simply the sum of $N^+$ independently chosen $p$'s following the distribution $f(p)$. We denote this by $\Delta k^+$, and do the same for the left-moving ones. 

Since $N^\pm$ follows the Poisson distribution (\ref{eq:Npmdist}), $\Delta k^\pm$ follows a compound Poisson exponential distribution. Its mean is, by the symmetry of $f(p)$ with respect to $p = 0$, 
\begin{equation}
    \langle \Delta k^\pm \rangle \, = \, \langle N^\pm \rangle \langle p^\pm \rangle \, = \, \pm \lambda(c \mp v)\Delta t \int^\infty_0 p f(p) dp \label{eq:DeltakpmDeltatavg}
\end{equation}
where $p^\pm$ denotes each half domain of $p$, and the variance is 
\begin{equation}
    \sigma^2_{\Delta k^\pm} \, = \, \langle N^\pm \rangle \langle (p^{\pm})^2 \rangle \, = \, \lambda(c \mp v)\Delta t \int^\infty_0 p^2 f(p) dp. \label{eq:DeltakpmDeltatvar}
\end{equation}

If $\Delta t$ is large enough that the average number of collision $\langle N^\pm \rangle$ is large, we can effectively regard that almost all of the actual $N^\pm$'s are large by Eq.~\eqref{eq:Npmdist}. Then, for each of them, the central limit theorem gives the distribution of $\Delta k^\pm$ to be well approximated by a Gaussian distribution. Consequently, the final $\Delta k^\pm$ is a weighted average of Gaussian random variables hence being Gaussian distributed. Then, as we already obtained its mean and variance in Eqs.~\eqref{eq:DeltakpmDeltatavg} and \eqref{eq:DeltakpmDeltatvar}, we have
\begin{eqnarray}
    \Delta k^\pm & \, \sim \, & \mathcal{N} \left( \pm \lambda(c \mp v)\Delta t \int^\infty_0 p f(p) dp \ , \ \lambda(c \mp v)\Delta t \int^\infty_0 p^2 f(p) dp  \right).
\end{eqnarray}
Finally, $\Delta k = \Delta k^+ + \Delta k^-$ gives Eq.~(\ref{eq:Deltakfinal}). 

\begin{figure} \centering
\includegraphics[width=0.49\linewidth]{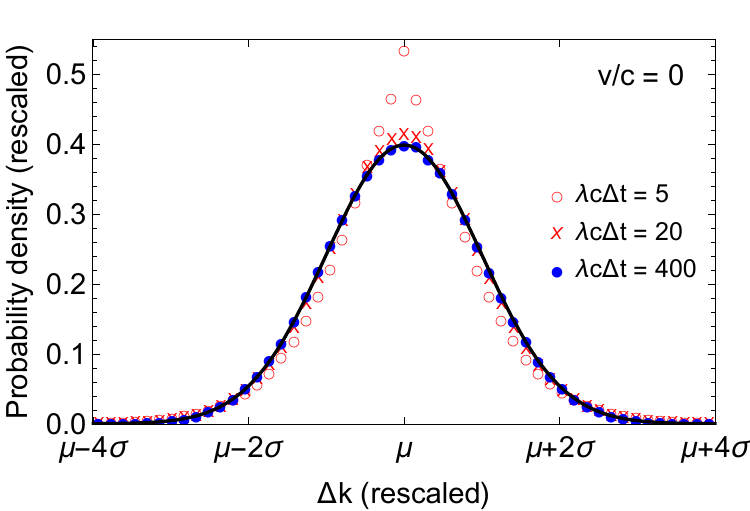}
\includegraphics[width=0.49\linewidth]{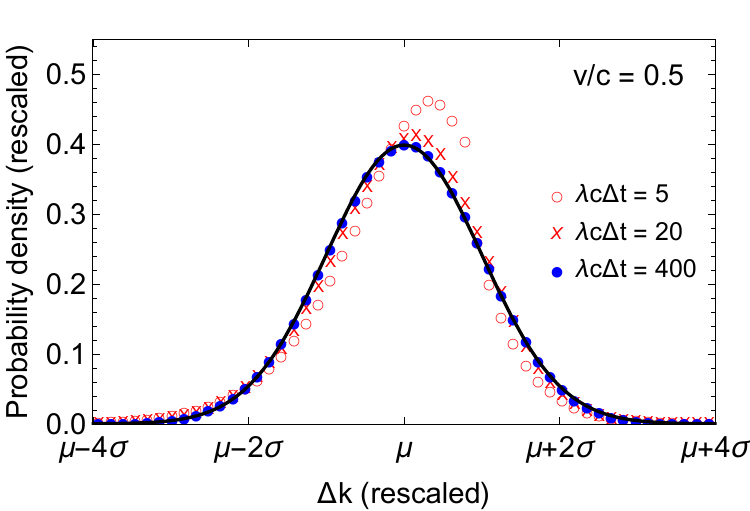}
\caption{Numerical simulation for the actual $\Delta k$ for $v = 0$ (left) and $v = 0.5 c$ (right), for three different $\Delta t$'s. The histograms obtained for $10^6$ trials for each case of $(v/c, \lambda c \Delta t)$ are converted to the probability density and overlayed with the one from the Gaussian distribution in Eq.~\eqref{eq:Deltakfinal} (black curve). The axes are rescaled for each case in order to have the same Gaussian curve for all of them. $\mu$ and $\sigma$ in the horizontal axes are also from Eq.~\eqref{eq:Deltakfinal}. 
} 
\label{fig:impulsesimulation}
\end{figure}

To track the actual distribution of $\Delta k$ at a finite $\Delta t$ and see how large $\Delta t$ is required for the Gaussianity, we ran numerical simulations for $v/c \leq 0.999$ and $\lambda c \Delta t \leq 3\times 10^4$. In Fig.~\ref{fig:impulsesimulation}, we show two example cases (see the caption for details). We see that the actual distribution of $\Delta k$ significantly differs from Gaussian for small $\lambda c \Delta t$'s, even with a sharp change of the probability density for the lowest $\lambda c \Delta t$ case for $v \neq 0$ due to the stepwise change in the collision rate in Eq.~\eqref{eq:dr}. However, the distribution well converges to the Gaussian in Eq.~\eqref{eq:Deltakfinal} for $\lambda c \Delta t \gtrsim \mathcal{O}(100)$. We checked that this conclusion holds true for all $v/c$'s chosen in the above range. Reverting the variables shows that
\begin{equation}
    \Delta t \, \gtrsim \, \mathcal{O}(100) \times \frac{H}{24\pi^2 M_P^2} \, \sim \, \frac{H^2}{M_P^2} \times \frac{1}{H} \, \equiv \, \Delta t_m
\end{equation}
is sufficient to have Gaussian distributed momentum change. Therefore, granted $H \ll M_P$, even a time scale much shorter than the Hubble time properly reproduces the quantum noise in Eq.~\eqref{eq:Langevin} or \eqref{eq:Langevink}. 

One possible caveat in the above derivation is that $v$ is assumed as a constant. If $V_\phi$ is not a linear function of $\phi$, $v$ can vary with time. From Eq.~(\ref{eq:Langevink}) with the limit in Eq.~\eqref{eq:cVp}, both the deterministic and the stochastic changes of $v$ during $\Delta t_m$ are restricted to be $\Delta v_m \lesssim HV''_\phi / M_P^2 m_\phi$. Such a variation of $v$ can induce a change in the peak position of the distribution of $\Delta k$ which may hinder the Gaussianity. So we require it to be much smaller than $\sigma_{\Delta k}$ for $\Delta t_m$, which is equivalent to $V''_\phi / M_P^2 \ll 1$. Since the 2nd potential slow-roll parameter is $\eta_V = V''_\phi / 3H^2$, the required condition is always satisfied as long as $\eta_V \ll 1$ and $H \ll M_P$, leaving the conclusion intact.


\section{Derivation with generic prefactors} \label{app:generalderiv}
In this appendix, we derive the results of the main text without presuming any numerical prefactors in relations. We start with minimal assumptions, the emergent particles with unspecified size in our space, the dual description of scalar fields, and classical mechanics in the abstract space. 

From the dual description, we let
\begin{equation}
    k \, = \, \mathfrak{C} \phi \label{eq:kgen}
\end{equation}
where $\mathfrak{C}$ is a dimensionless constant and
\begin{equation}
    E_k \, = \, \mathfrak{V} V_\phi \, = \, \mathfrak{V} \int V'_\phi d\phi \label{eq:Ekgen}
\end{equation}
where $\mathfrak{V}$ is a constant with a dimension of volume, which is the volume of an emergent particle in our 3D space. With the work-energy theorem \eqref{eq:WEtheorem} and Eq.~\eqref{eq:kgen}, Eq.~\eqref{eq:Ekgen} gives the velocity 
\begin{equation}
    v \, = \, \frac{\mathfrak{V}}{\mathfrak{C}} V'_\phi. \label{eq:vgen}
\end{equation}

To obtain the expression for the mass $M$ without assuming the same volume factor, we use the Newtonian kinetic energy formula near $\phi = 0$ with $V_\phi \simeq \frac{1}{2} m_\phi^2 \phi^2$, giving the counterpart of Eq.~\eqref{eq:Eknearmin} as
\begin{equation}
    E_k \, \simeq \, \frac{1}{2M} \left( \mathfrak{C} \phi \right)^2 \, \simeq \, \frac{1}{2}M \left( \frac{\mathfrak{V}}{\mathfrak{C}} m_\phi^2 \phi \right)^2, \label{eq:Eknearmingen}
\end{equation}
where we used Eq.~\eqref{eq:vgen} in the place of $v$. This gives
\begin{equation}
    M \, = \, \frac{\mathfrak{C}^2}{\mathfrak{V} m_\phi^2}. \label{eq:Mgen}
\end{equation}

Then we make an additional assumption that if $E_k$ has a relation with $V_\phi$ in Eq.~\eqref{eq:Ekgen}, $M$ should follow the same relation to $V_0$ as
\begin{equation}
    M \, = \, \mathfrak{V} \, 3 M_P^2 H^2. \label{eq:MgenV}
\end{equation}
Eqs.~\eqref{eq:Mgen} and \eqref{eq:MgenV} then fixes the relation between $\mathfrak{C}$ and $\mathfrak{V}$ as
\begin{equation}
    \mathfrak{C} \, = \, \mathfrak{V} \, \sqrt{3} M_P H m_\phi.
\end{equation}
Now the only remaining undetermined scale in the formalism is $\mathfrak{V}$. We see that if $\mathfrak{V} = \frac{4\pi}{3H^3}$, we have $\mathfrak{C} = \frac{4\pi M_p m_\phi}{\sqrt{3} H^2}$ and Eqs.~\eqref{eq:Ek}, \eqref{eq:M}, \eqref{eq:k}, and \eqref{eq:v} are recovered. 

To maintain generality, we let $\mathfrak{V} = \alpha_1 \, \frac{4\pi}{3H^3}$ with dimensionless free parameter $\alpha_1$. Then, in compared to the expressions in Eqs.~\eqref{eq:Ek}, \eqref{eq:M}, \eqref{eq:k}, and \eqref{eq:v}, $E_k$, $M$, and $k$ get an additional factor of $\alpha_1$ while $v$ remains intact. This gives the following modifications. First, the equilibrium temperature extracted from Eq.~\eqref{eq:rhophiEquil} becomes $\alpha_1 T_{\rm dS}$. Second, the r.h.s. of Eq.~\eqref{eq:Langevink} gets an additional factor of $\alpha_1$. And third, the r.h.s. of Eq.~\eqref{eq:dEdtHub} also gets the same additional factor. 

We move on to the same heat bath model with the same parameters, $c$, $T$, and $\lambda$. But to be general, we modify the energy postulation by equating the energy loss rate in Eq.~\eqref{eq:DeltaEfinal} to a constant $\alpha_2$ multiple of the required energy gain rate in Eq.~\eqref{eq:dEdtHub}. Then, the last equalities of Eqs.~\eqref{eq:Deltakidentify} -- \eqref{eq:DeltaEidentify} become
\begin{eqnarray}
    \frac{\lambda v}{\beta c} & \, = \, & \alpha_1 \, \frac{4\pi M_P^2 m_\phi^2}{3H^2}v \label{eq:Deltakidentifygen} \\
    \frac{2 \lambda}{\beta^2 c} & \, = \, & \alpha_1^2\, \frac{4M_P^2 m_\phi^2}{3H} \label{eq:sigmasqDeltakidentifygen} \\
    \frac{\lambda c}{\beta} & \, = \, & \alpha_1 \alpha_2 \, 12\pi M_P^2 \label{eq:DeltaEidentifygen}
\end{eqnarray}
resulting in 
\begin{eqnarray}
    c & \, = \, & \sqrt{\alpha_2} \, \frac{3H}{m_\phi} \\
    T & \, = \, & \alpha_1 T_{\rm dS} \\
    \lambda & \, = \, & \sqrt{\alpha_2} \, \frac{8 \pi^2 M_P^2 m_\phi}{H^2}.
\end{eqnarray}

Therefore, we conclude with the following. The only free scale of the formalism is the volume of an emergent particle in our 3D space. For the heat bath model, we may allow an additional freedom to the energy postulation, at the cost of losing energy conservation. Then we observe the three scaling behaviors. First, the equality between the bath temperature and the equilibrium temperature from the Fokker-Planck equation always holds, making the model physically consistent. Second, that temperature is proportional to the volume of an emergent particle but not affected by the energy postulation, and $T = T_{\rm dS}$ holds only when that volume equals the Hubble volume. Lastly, the speed of light in the abstract space is only affected by the energy postulation but not by the volume choice. We discussed the physical implications of these results in the main text.


\section{Maxwell-Boltzmann distribution from quantum statistics} \label{app:mixedptl}
In this appendix, we explicitly show that the Maxwell-Boltzmann statistics can be obtained for quantum emergent particles by adding the auxiliary space.

First, consider a density operator 
\begin{equation}
    \hat{\rho} \, = \, \frac{1}{Z} e^{-\beta (\hat{H} - \mu \hat{N})} \label{eq:rho}
\end{equation}
with the partition function
\begin{equation}
    Z \, = \, \text{Tr}\left[e^{-\beta (\hat{H} - \mu \hat{N})}\right] \label{eq:partfunc}
\end{equation}
where $\hat{H}$ is the Hamiltonian and $\hat{N}$ is the total number operator for emergent particles. The trace is taken over all the possible states. 

We now assume that the emergent particles are approximately free in the abstract space despite interacting with the bath particles. This approximation would hold when the interaction is close-ranged or the strength is weak, or near the end of the dS stage where the bath particles are depleted. Otherwise, corrections to the resultant equilibrium distribution would arise. We leave any deviations from this situation to future studies and focus on the free case here. Then, the Hamiltonian and the total number operator can be written as
\begin{equation}
    \hat{H} \, = \, \sum_{k, \vec{q}} \omega_{k, \vec{q}} \hat{N}_{k, \vec{q}} \ \text{,} \quad \hat{N} \, = \, \sum_{k, \vec{q}} \hat{N}_{k, \vec{q}} \label{eq:HN}
\end{equation}
where
\begin{equation}
    \hat{N}_{k, \vec{q}} \, = \, \hat{o}_{k, \vec{q}}^\dagger \, \hat{o}_{k, \vec{q}} \, = \, (\hat{a}_k^\dagger \, \hat{a}_k) \otimes (\hat{b}_{\vec{q}}^\dagger \, \hat{b}_{\vec{q}}) . \label{eq:Nkq}
\end{equation}
$\omega_{k, \vec{q}}$ is the energy per emergent particle having $k$ and $\vec{q}$, and we assume that it is given as a sum of $E_k$ in Eq.~(\ref{eq:Ek}) and an unspecified contribution from $\vec{q}$ as
\begin{equation}
    \omega_{k, \vec{q}} \, = \, E_k + \epsilon_{\vec{q}}.  \label{eq:omegakq}
\end{equation}

Being diagonalized in the number basis, we evaluate the partition function in Eq.~(\ref{eq:partfunc}). Each state expressed in the number basis is
\begin{equation}
    |\psi_{\{n\}} \rangle \, = \, |n_{k_1, \vec{q}_1}\rangle \, |n_{k_1, \vec{q}_2}\rangle \, \cdots \, |n_{k_2, \vec{q}_1}\rangle \, |n_{k_2, \vec{q}_2}\rangle \, \cdots,
\end{equation}
so
\begin{equation}
    \langle \psi_{\{n\}} | e^{-\beta (\hat{H} - \mu \hat{N})} | \psi_{\{n\}} \rangle \, = \, \prod_{k, \vec{q}} e^{-\beta (\omega_{k, \vec{q}} - \mu) \, n_{k, \vec{q}}}. \label{eq:psirhopsi}
\end{equation}
To obtain the partition function, we sum Eq.~(\ref{eq:psirhopsi}) for all possible combinations of particle numbers $\{n\}$. Due to the exclusiveness on $\vec{q}$, each $n_{k, \vec{q}}$ can be either 0 or 1, and at most only one of $n_{k, \vec{q}}$'s with the same $\vec{q}$ (but with different $k$'s) can be 1. Had none of the constraints been applied, the ordinary Bose-Einstein statistics would appear with each $(k, \vec{q})$ state being all independent. Had only the first constraint been applied, the ordinary Fermi-Dirac statistics would appear with each $(k, \vec{q})$ state being all independent. With the two constraints together, the states with the same $\vec{q}$ are considered as one group that can hold up to 1 particle. When projected onto the abstract space, these groups are distinguishable by $\vec{q}$ and act as distinguishable particles. 

Expanding the product in the r.h.s. of Eq.~(\ref{eq:psirhopsi}) gives
\begin{eqnarray}
    && \langle \psi_{\{n\}} | e^{-\beta (\hat{H} - \mu \hat{N})} | \psi_{\{n\}} \rangle \nonumber \\
    &&\, \, = \, \left[ e^{-\beta(\omega_{k_1, \vec{q}_1} - \mu) n_{k_1, \vec{q}_1} } \times e^{-\beta(\omega_{k_2, \vec{q}_1} - \mu) n_{k_2, \vec{q}_1} } \times e^{-\beta(\omega_{k_3, \vec{q}_1} - \mu) n_{k_3, \vec{q}_1} } \times \cdots \right] \nonumber \\
    &&\, \phantom{\, = \,} \times \left[ e^{-\beta(\omega_{k_1, \vec{q}_2} - \mu) n_{k_1, \vec{q}_2} } \times e^{-\beta(\omega_{k_2, \vec{q}_2} - \mu) n_{k_2, \vec{q}_2} } \times e^{-\beta(\omega_{k_3, \vec{q}_2} - \mu) n_{k_3, \vec{q}_2} } \times \cdots \right] \nonumber \\
    &&\, \phantom{\, = \,} \times \left[ e^{-\beta(\omega_{k_1, \vec{q}_3} - \mu) n_{k_1, \vec{q}_3} } \times e^{-\beta(\omega_{k_2, \vec{q}_3} - \mu) n_{k_2, \vec{q}_3} } \times e^{-\beta(\omega_{k_3, \vec{q}_3} - \mu) n_{k_3, \vec{q}_3} } \times \cdots \right] \nonumber \\
    &&\, \phantom{\, = \,} \times \cdots
\end{eqnarray}
where each square bracket is the product of terms with different $k$'s but with the same $\vec{q}$. Inside each bracket, at most only one term can survive with $n_{k, \vec{q}} = 1$, and others are all unity with $n_{k, \vec{q}} = 0$. So after summing over all the possible valid combinations, we can reorganize the sum of products into a product of sums as
\begin{eqnarray}
    Z & \, = \, & \sum_{\{n\}} \langle \psi_{\{n\}} | e^{-\beta (\hat{H} - \mu \hat{N})} | \psi_{\{n\}} \rangle \nonumber \\
    & \, = \, & \left[1 + e^{-\beta(\omega_{k_1, \vec{q}_1} - \mu)} + e^{-\beta(\omega_{k_2, \vec{q}_1} - \mu)} + e^{-\beta(\omega_{k_3, \vec{q}_1} - \mu)} + \cdots \right] \nonumber \\
    && \times \left[1 + e^{-\beta(\omega_{k_1, \vec{q}_2} - \mu)} + e^{-\beta(\omega_{k_2, \vec{q}_2} - \mu)} + e^{-\beta(\omega_{k_3, \vec{q}_2} - \mu)} + \cdots \right] \nonumber \\
    && \times \left[1 + e^{-\beta(\omega_{k_1, \vec{q}_3} - \mu)} + e^{-\beta(\omega_{k_2, \vec{q}_3} - \mu)} + e^{-\beta(\omega_{k_3, \vec{q}_3} - \mu)} + \cdots \right] \nonumber \\
    && \times \cdots \nonumber \\
    & \, = \, & \prod_{\vec{q}} \left[1 + \sum_k e^{-\beta (\omega_{k, \vec{q}} \, - \mu)} \right]. \label{eq:Zfinal}
\end{eqnarray}

To obtain the average number of particles in each state $\langle \hat{N}_{k, \vec{q}} \rangle$, we go back to Eq.~(\ref{eq:rho}) and use Eq.~(\ref{eq:HN}). We have 
\begin{eqnarray}
    \langle \hat{N}_{k, \vec{q}} \rangle & \, = \, & -\frac{1}{\beta Z} \frac{\partial Z}{\partial \, \omega_{k, \vec{q}}} \, = \, \frac{e^{-\beta (\omega_{k, \vec{q}} - \mu)} \times \prod_{\vec{q}' \neq \vec{q}} \left[1 + \sum_{k'} e^{-\beta (\omega_{k', \vec{q}'} \, - \mu)} \right]}{\prod_{\vec{q}'} \left[1 + \sum_{k'} e^{-\beta (\omega_{k', \vec{q}'} \, - \mu)} \right]} \nonumber \\
    & \, = \, & \frac{e^{\beta \mu} e^{-\beta \epsilon_{\vec{q}}} e^{-\beta E_k} }{1 + e^{\beta \mu} e^{-\beta \epsilon_{\vec{q}}} \sum_{k'} e^{-\beta E_{k'}}}
\end{eqnarray}
where the last equality uses Eq.~(\ref{eq:omegakq}). Since the denominator does not depend on $k$, the distribution on $k$ for every $\vec{q}$ is proportional to 
\begin{equation}
    \langle \hat{N}_{k, \vec{q}} \rangle \, \propto \, e^{-\beta E_k} \label{eq:Nkqavgfinal}
\end{equation}
and after summing over all $\vec{q}'s$, the same proportionality applies to the entire emergent particles as desired.

\bibliographystyle{JHEP}
\bibliography{references.bib}

\end{document}